\documentclass[showpacs,aps]{revtex4}
\begin{document}

\title{Transitions involving conical magnetic phases in a model
with bilinear and biquadratic interactions.}
\author{Vittorio Massidda}

\affiliation{Departamento de F\'{\i}sica, Comisi\'on Nacional de
Energ\'{\i}a At\'omica, \\ 
Avda. General Paz 1499, C.P.1650 San Mart\'{\i}n, Prov. de Buenos Aires,\\ Argentina \\
Fax: 54 11 6772-7121,  \hspace{2 mm}  E-mail: massidda@cnea.gov.ar}

\begin{abstract}

 In a previous work a model was proposed for the phase transitions of crystals with localized
 magnetic moments which at low temperature have a "conical" arrangement
 that at higher $T$ transforms into a more symmetrical structure
 (depending on the compound) before becoming totally disordered.
 The model assumes bilinear and biquadratic interactions between magnetic
 moments up to the fifth neighbours, and for any given $T$ the structure with
 the least free energy is obtained by a mean-field approximation (MFA). The interaction
 constants are derived from {\em ab initio} energy calculations.

 In this work we improve upon that model modifying the MFA in such a way that a continuous (instead of
 discontinuous) spectrum of excited states is available to the system.
 In the previous work, which dealt with LaMn$_2$Ge$_2$ and LaMn$_2$Si$_2$, we found that
 transitions to different structures can be obtained for increasing $T$, in good qualitative agreement
 with experiment. The critical temperatures, however, were exaggerately high.
 With the new MFA we obtain essentially the same behaviour concerning the phase transitions,
 and critical temperatures much closer to the experimental ones.
 
\end{abstract}


 \maketitle

\thispagestyle{empty}


\section{Introduction}

 Localized magnetic moments in several compounds  of the family
 RT$_2$X$_2$ (R = rare earth, T = transition metal, X = Ge, Si) are found
 to have "conical" ground state (GS) configurations (see eq.1) and to undergo
 transitions to configurations with higher symmetry (helical, canted, collinear)
 as temperature $T$ increases \cite{1}. Among the many works on this subject we can
 cite refs.2-6 (see also references therein and in ref.1).
 In this work we consider LaMn$_2$Ge$_2$ and LaMn$_2$Si$_2$ \cite{2},\cite{3},\cite{4}.
 In such compounds only the Mn atoms bear a localized magnetic moment, so that we
 consider only the Mn sublattice. We take it as a base-centered tetragonal lattice,
 whose lattice constants are given, in terms of the crystallographic constants
 $a$ and $c$, by $ a\prime \equiv a$  and $ c\prime \equiv c/2 $, and the lattice sites by
 $\vec{R}_{hkl} \equiv h a\prime \vec{e}_x + k a\prime \vec{e}_y + l c\prime \vec{e}_z $,
 where $l$ is an integer and $j,k$ are both integers or half-integers.
 In the following we will use subindices $i$
 and $j$ as short notations for sets $(h,k,l)$.
 We work with the total atomic angular momentum (in units of $\hbar$),
 which we call spin and denote $\vec{S}$, as in the spin Hamiltonian
 formalism. The corresponding magnetic moment is
 $ \vec{\mu} \equiv \mu_B \vec{S} $ ( $\mu_B$ is Bohr's magneton).

 The magnetic structure can be characterized by a pair
 of polar angles $\theta$ and $\alpha$, such that at any site
 $\vec{R}_{hkl} \equiv h a'\vec{e}_x + k a'\vec{e}_y + l c'\vec{e}_z $
 the average value of the local spin is given by
\begin{equation}
 \overline{\vec{S}_{hkl} }= \bar{S} [sin(\theta)cos(l \alpha + \xi_{hk}) \vec{e_x}
 + sin(\theta)sin(l \alpha + \xi_{hk}) \vec{e_y} + cos(\theta) \vec{e_z} ] ,
\end{equation}
 where $\xi_{hk}=0$ or $ \pi$ for $h,k$ integers or half-integers respectively
 and the bars indicate the statistical averages (see also eqs. 22-24).
We can also say that the polar angles ($\theta_i$, $\phi_i$)
 characterizing the direction of the average spin at a site
 $i \equiv (h,k,l)$ are given by
 $\theta_i = \theta$ and $\phi_i = (l \alpha + \xi_{hk})$.
 For general values of $\theta$ and $\alpha$ (i.e. values
 different from 0, $\frac{1}{2}\pi$, $\pi$ and $\frac{3}{2}\pi$)
 this gives what is customarily called a conical structure (even
 though perhaps the term "conical" {\em tout court} should be limited
 to the case $ \xi_{hk} = 0 $ $\forall$ $ h,k$).
 Particular cases of the conical structure are the helical ($ \theta = \frac{1}{2}\pi$),
 the canted ( $\alpha =  0, \frac{1}{2}\pi$, $\pi$ or $\frac{3}{2}\pi$") and the collinear
 (all the spins are parallel or anti-parallel to each other) structures.
 In the following we will call "general conical structure" a conical structure in which
 $ \theta $ and $ \alpha $ can take any values, including these particular values.
 
 As $T$ increases the spins get progressively disordered (so
 that their average values decrease), while angles $\theta$ and $\alpha$
 vary, until eventually taking values corresponding to some of the above-mentioned
 "particular cases" so that a more symmetrical structure is obtained.
 In a previous work\cite{7} a model was proposed assuming that the
 localized magnetic moments are subjected to bilinear and biquadratic
 interactions up to fifth neighbours (and possibly to an anisotropy
 field). The interaction constants were obtained 
 from {\em ab initio} calculations of the energies of different structures.

  For any given $T$ the structure was obtained by finding,in a mean-field
 approximation (MFA), the $\theta$ and $\alpha$
 angles, the modulus of the average spins, and the average of $ m^2 $
 ($ m = $ total angular momentum quantum number) for which the free energy
 was minimal. This model could qualitatively 
 account in a reasonable way for the mentioned behaviours, but it
 gave exaggerately high values for the transition temperatures $T_c$'s
 between the different configurations (see sections IV,V).
 In this paper we modify the MFA in such a way that
 reasonable values of the $T_c$'s are obtained.

\section{The model}

A common approach to a phenomenological study of a crystal with localized 
magnetic moments is the employ of the bilinear Heisenberg Hamiltonian:
\begin{eqnarray}
\hat{H}_J = - \frac{1}{2}
 \sum_{i,j}{ }^{'} J_{ij} \hat{\vec{S}_i} \cdot \hat{\vec{S}_j}
\end{eqnarray}
(the prime in the summation means $i \neq j$).

 The ground state for such a Hamiltonian is a general type of
 helical structure \cite{8}, so that in order to obtain a
 conical structure some other interaction must be added to it.
 The addition of a biquadratic term to the Heisenberg bilinear interaction
 was shown to explain another type of phase transitions in a RT$_2$X$_2$ compound, namely
 U Ni$_2$Si$_2$ \cite{9}.
 In ref.7 it was found that a Hamiltonian with bilinear and 
 biquadratic interactions both up to the fifth neighbouring sites
 gives an acceptable fitting of {\em ab initio} total energy values
 for several configurations in which the moduli of the spins are all equal to $S$.
 The anisotropy field was not required, but we included the quadratic term
 in order to study its possible effects (a $\cos^2(\theta)$ term does anyway
 appear in the bilinear interaction).

 Accordingly, we assume the Hamiltonian 

 \begin{eqnarray}
  \hat{H} = \hat{H}_J + \hat{H}_B + \hat{H}_D
 \end{eqnarray}

 with
 \begin{eqnarray}
  \hat{H}_B \equiv - \frac{1}{2} \sum_{i,j}{ }^{'}
 B_{ij} [ \hat{\vec{S}_i} \cdot \hat{\vec{S}_j} ]^2
 \end{eqnarray}
 and

 \begin{eqnarray}
 \hat{H}_D \equiv D_2 \sum_{i} \hat{S}_{i;z}^2
 \end{eqnarray}

 where the $J_{ij}$ and $B_{ij}$ are non-zero for the five nearest sets of
 neighbours (which, for the reference site (0,0,0), are:
 $(\pm\frac{1}{2} \pm\frac{1}{2},0)$; $(\pm 1,0,0)$ and $(0,\pm 1,0)$; 
 $(0,0,\pm 1)$; $(\pm 1,\pm 1,0)$; $(\pm\frac{1}{2}, \pm\frac{1}{2},\pm1)$ ).

 Let us denote $ \mid m_i >$ the eigenstate of a spin at site $i$
 with eigenvalue $m_i$ in the direction given by the polar angles
 ($ \theta_i, \phi_i$).

 The interaction energy between the spins at sites $i$ and $j$ (i.e.

 $  < m_i \! \mid  < m_j  \! \mid (\hat{H}_J + \hat{H}_B) \mid m_i > \mid m_j >$)  is
 \begin{eqnarray}
 U_{JB;ij}(m_i,m_j) \equiv U_{J;ij}(m_i,m_j) + U_{B;ij}(m_i,m_j) ,
 \end{eqnarray}

 with

\begin{eqnarray}
U_{J;ij}(m_i,m_j) = - \frac{1}{2} J_{ij} m_i m_j p_{ij}
 \end{eqnarray}

and

\begin{eqnarray}
 U_{B;ij}(m_i,m_j) =
 - \frac{1}{2} B_{ij} \left[ \frac{1}{4} [\!S(S\!+1) -3m_i^2][\!S(\!S+\!1) -3m_j^2]
 (p_{ij}^2 + 1) + S(S+1)(m_i^2 + m_j^2) - 3 m_i^2 m_j^2 -
 \frac{1}{2} m_i m_j p_{ij} \right]
 \end{eqnarray}

where $p_{ij}$ is the cosine of the angle between the spins
at sites $i$ and $j$.
The matrix element of $\hat{H}_B$ was calculated by rotating
 the spin operators \cite{10}.

In a general conical structure all sites are energetically equivalent to each
 other, so that neglecting boundary effects the total energy is equal
 to $N U_s$, where $N$ is the total number of sites and $U_s$ is the
 energy per site. As reference site we take $(h = k = l = 0)$, also
 denoted $i=0$. We have

\begin{eqnarray}
U_s = \sum_{j=1}^{N-1} U_{JB;0j}(m_0,m_j) +
 \frac{1}{2} D_2 \left[ \left[S(S+1) - m_0^2\right] +
  \left[3m_0^2 - S(S+1) \right] \cos^2(\theta) \right].
\end{eqnarray}

 We shall denote respectively $J_\nu$ and $B_\nu$ the constants $J_{ij}$ and $B_{ij}$ for
 sites $i,j$ that are $\nu^{{\rm th} }$ neighbours of each other.

 In ref.7 eq.(9) with
 $m_0 = m_j = S$  was used to determine the parameters ($J_\nu, B_\nu, D_2$ of the Hamiltonian.
 The energy dependence on $\theta$ and $\alpha$ was written as
\begin{eqnarray}
U_{T=0}(\theta, \alpha) =   X_1 + X_2 \cos^2 \theta +
 X_3 \sin^2 \theta \cos \alpha  + X_4 \cos^4 \theta +
 X_5 (\cos^2 \theta + \sin^2 \theta \cos \alpha)^2 +
 X_6 (\cos^2 \theta - \sin^2 \theta \cos \alpha)^2 ,
\end{eqnarray}
 where the $X_n$'s are constants.

 In principle the $X_n$'s can be obtained by fitting the results of
 {\em ab initio} total energy calculations for several configurations
 with $m_0 = m_j = S$. Such energies were obtained by
 using the FLEUR code, as described in ref.7.
 If this fitting is done for six configurations, i.e. for six pairs
 ($\theta,\alpha$), a linear system is obtained, which can be solved
 for the $X_n$.
 We did this for several sets of six configurations, so as to check 
 the consistency of the formalism (i.e, that in each case the same set of 
 $X_n$ is obtained with a reasonable accuracy). 
 We found that this consistency is achieved if these configurations 
 are neither too close to each other nor too far from the GS.
 In the former case the determinants appearing in the numerical 
 solution of the system are nearly zero (they are sums of terms
 which nearly cancel out each other);
 in the latter case it is possible that if one forces the electrons
 associated with the magnetism to occupy high-energy states, one cannot
 neglect the additional energy due to changes of the other electronic
 states. In both cases the error is large.

 The $X_n$'s to be adopted in this work for LaMn$_2$Ge$_2$ and
 LaMn$_2$Si$_2$ are the averages of the $X_n$'s
 of the selfconsistent sets.

 Having obtained the $X_n$'s, we must face the problem of determining
 the eleven parameters appearing in the Hamiltonian.

 The part of the magnetic energy not depending on $\theta$ and $\alpha$
 (i.e. $X_1$) is not the same as the analogous part of the total energy
 (as a matter of fact, it is several orders of magnitude smaller),
 so that it cannot be singled out in the {\em ab initio} values.
 Therefore, $X_1$ is of no use for the determination of the parameters.
 As a consequence, $J_2$, $J_4$, $B_2$ and $B_4$ cannot
 be determined because they multiply terms with $p_{02}$ or $p_{04}$,
 which in a conical structure are constant, being equal to 1.
 In order to express the other seven parameters
 in terms of the five constants $X_n$ ($n$=2,$\ldots$,6) we compare eq.(10)
 with the energy per site (eq.9 with (6-8)) for
 $ m_i = m_j = S $ $\forall$ $i,j$. In doing so we must take into account
 that any site has four first, second and third neighbours, two fourth
 neighbours and eight fifth neighbours. We obtain the system
 
 \begin{mathletters}
 \begin{eqnarray}
 X_2 &=& -\frac{1}{2}S^2 [ 4(2 J_1 - B_1) + (2 J_3 - B_3) +
 4 (2 J_{5} - B_5)] + 8 S^2 (S-\frac{1}{2})^2 B_1 +
 S (S - \frac{1}{2}) D_2   \nonumber \\
 X_3 &=& - \frac{1}{2} S^2 ( 2J_3 - B_3 - 8 J_5 + 4 B_5)   \nonumber \\
 X_4 &=& - 8 S^2 (S-\frac{1}{2})^2 B_1 ]   \\
 X_5 &=& - S^2 (S-\frac{1}{2})^2 B_3   \nonumber \\
 X_6 &=& - 4 S^2 (S-\frac{1}{2})^2 B_5   \nonumber
 \end{eqnarray}
 \end{mathletters}

From here we obtain directly $B_1$, $B_3$ and $B_5$ , after
which we are left with two equations for the unknowns
 $J_1$, $J_3$, $J_5$ and $D_2$, two of which must be chosen
 arbitrarily (as well as $J_2$, $J_4$, $B_2$ and $B_4$).
 Concerning $D_2$, we shall give it a small value, while for the other
 three unknowns we think that the less arbitrary procedure is that of
 choosing $J_5$ in such a way that it and the values of $J_1$ and
 $J_3$ calculated in terms of it and of $D_2$ have a smooth variation
 with the interatomic distance, and are, in principle, about an order
 of magnitude greater than the corresponding $B_j$. An analogous
 criterion shall be followed for $J_2$, $J_4$, $B_2$ and $B_4$.

  Setting $ \frac{1}{2} S^2 (S-\frac{1}{2})^2 \equiv C_2 $ we have

\begin{mathletters}
\begin{eqnarray}
 B_1 &=& - \frac{1}{16} X_4/C_2  \\
 B_3 &=& - \frac{1}{2} X_5/C_2    \\
 B_5 &=& - \frac{1}{8}X_6/C_2    \\
 J_3 &=& - \frac{1}{2} X_3/S^2 + \frac{1}{2} B_3 +  J_5 - 2 B_5   \\
J_1 &=& \frac{1}{4} { - X_2/S^2 + [8 (S-\frac{1}{2})^2 + 2] B_1 
 - J_3 + \frac{1}{2} B_3 - 4 J_5 + 2 B_5 +
 [(S - \frac{1}{2})/S] D_2 }   
\end{eqnarray}
\end{mathletters}
\section{Calculation of the free energy}

 In this work we must find the magnetic structure of the LaMn$_2$X$_2$
 compounds (which we know from experiment to be a general conical structure)
 at any temperature $T$. Therefore we must determine which is for the different
 sets of values of the interaction constants the conical
 structure that minimizes the free energy.
 In a general state of the system the spin at each site $i$ is in a
 state $ \mid m_i>$, so the system is characterized by the $3N$ quantities $m_i$, $\theta_i$, $\phi_i$
 ($i=1,\ldots,N$).
 
 For $T > 0$ the state of the system will be that superposition of
 the above-defined states which minimizes the free energy
 $F = -k_B T \log Z $, i.e.which maximizes the partition function.
 The latter is given by
 
 \begin{eqnarray}
 Z = \sum_{m_0,\ldots,m_{N-1} = -S}^S \: \: \int d\Omega_0 ... \int d\Omega_{N-1}
 \exp[ - \beta U(m_i,\theta_i,\phi_i)] ,
 \end{eqnarray}

where $ \beta \equiv = 1/ kT $, $d\Omega_i \equiv \sin \theta_i d\theta_i d\phi_i $ and each
integration is carried out over all the space directions.

Taking advantage of the fact that all the sites are physically equivalent
 to each other we define the partition function per site $Z_s$ through
 $Z \equiv Z_s^N$.

 We calculate $Z_s$ for the reference site $i=0$. To do this we must make a MFA, i.e we must
 approximate the energies of the states at that site in the field of the spins at the other sites
 (which will be referred to as "source sites")
 by replacing each of the latter by some average state. This is done as follows.
 The $m_j$, $\theta_j$ and $\alpha_j$ at each source site $j$ are taken according to eq.(1),
 where $\bar{S}$, $\theta$ and $\alpha$ are unknowns. Furthermore, as
 the contribution of a $j$ site to the energy of the reference spin
 depends also (due to the biquadratic interaction) on $m_j^2$ 
 we must assign to each source site a fourth variable, $\overline{m^2}$.
We have

\begin{eqnarray}
 Z_{ {\rm s,MFA} }(\theta,\alpha,\bar{S},\overline{m^2}) =
  \sum_{m_0 = -S}^S m_0 \int d\Omega_0
exp[ - \beta U(m_0,\theta_0,\phi_0;\theta,\alpha,\bar{S},\overline{m^2})] .
\end{eqnarray}

The energy in the integrand contains the contributions of all the source sites,
and consists of an interaction term and an anisotropy term (eqs.7,8,9).
 The latter, being a local term, requires no approximation.

 For the interaction energy associated with the reference site, whose exact value is
 
\begin{eqnarray}
 \sum_{j=1}^{N-1} U_{JB;0j}(m_0,m_j;p_{0j}) 
 \end{eqnarray}

 (where $ U_{JB;0j}(m_0,m_j;p_{0j}) $ is given by eq.(6) with (7,8) taking into account
 that the spin at the $i=0$ site has the general orientation ($\theta_0,\phi_0$) ),
there are several possible MFA expressions,
 and it is not clear which one is the best in any particular case (see section VI).

 Now we must find, for any given $T$, the values of 
 $\theta$, $\alpha$, $\bar{S}$, and $\overline{m^2}$ characterizing the state
 of minimal free energy, i.e. of maximal $Z$. For a conical
 structure the extremum conditions are

 \begin{mathletters}
 \begin{eqnarray}
 \partial Z / \partial \theta = 0  \\
 \partial Z / \partial \alpha = 0  .
 \end{eqnarray}
 \end{mathletters}

 On the other hand, both $\bar{S}$ and $\overline{m^2}$ must satisfy a
 self-consistence condition, i.e. they must be equal to the respective
 average values they give rise to at the reference site.
 $\bar{S}$ is the modulus of
 \begin{eqnarray}
 \bar{\vec{S}} = Z_s^{-1} \sum_{m_0 = -S}^S m_0 \int d\Omega_0
 [\sin \theta_0 \cos \phi_0 \,\vec{e}_x  + \sin \theta_0 \sin \phi_0 \,\vec{e}_y
  + \cos \theta_0 \,\vec{e}_z ]
 \exp[ - \beta U(m_0,\theta_0,\phi_0;\theta,\alpha,\bar{S},\overline{m^2})] ,
 \end{eqnarray}
 i.e. is given by
 \begin{eqnarray}
 \bar{S} = [\overline{S_x}^2 + \overline{S_z}^2]^{\frac{1}{2}}
 \end{eqnarray}
 (due to the symmetry of the structure, the $y$ component vanishes),
 while

\begin{eqnarray}
 \overline{m^2} = Z_s^{-1} \sum_{m_0 = -S}^S \int d\Omega_0 m_{0}^2
 \exp[ - \beta U(m_0,\theta_0,\phi_0;\theta,\alpha,\bar{S},\overline{m^2})] .
 \end{eqnarray}

 So, we must solve the system of equations (20), (21), (23) and (24) 
 \footnote{To speed up the calculation and improve its accuracy, the integrations
 over $\phi_0$ coming from eq.(12) are carried out analytically, the
 results being series of the modified Bessel functions $I_n(x)$ \cite{11}.}.

 For the more symmetrical structures the number of equations is reduced:
 for the helical structure we set $\theta = \frac{1}{2}\pi$ and drop eq.(16a);
 similarly, for the canted structure, we set
 $\alpha = \pi$, and instead of (16a) we drop (16b).
 Finally, in the collinear structure we have $\theta = \frac{1}{2}\pi$,
 $\alpha = \pi$, and the equation system is reduced to (14) and (15).

 \section{Numerical calculations. LaMn$_2$Ge$_2$}

 The interaction constants $J_{ij}, B_{ij}$ and $D_2$ are obtained from
 the $X_n$'s used in ref.7, taking into account that the latter were
 evaluated for a cell with four sites, i.e. must be divided by 4.
 According to the {\em ab initio} calculations of ref.7 and to the
 experimental results of ref.2 (where the helical and FM components of
 the Mn magnetic moment are found to be approximately $2.6\mu_{\rm B}$
 and $1.61\mu_{\rm B} $ respectively), and assuming $L=0$ because of
 quenching, we take $S = \frac{3}{2}$.

 As we said above, only $B_1$, $B_3$ and $B_5$ can be derived from the
 {\em ab initio} energies.  It turns out that these $B_{\nu}$'s are
 negative, and their absolute values decrease smoothly for increasing
 interatomic distance.

 For the other constants we proceed as explained in section 2:
 we choose first $J_5$ and $D_2$, thereby determining
 $J_1$ and $J_3$, and then, without affecting the other constants, we
 choose $J_2$, $J_4$, $B_2$ and $B_4$.
 We can get a set of $J_\nu$'s about an order of magnitude greater than
 the $B_\nu$'s only if $J_1$ and $J_3$ have different signs. This is a
 quite reasonable condition, being a property of the RKKY interaction
 \footnote{Of course, some $J_{\nu}$'s can be near to zero, in which case it can be
 smaller in modulus than the corresponding $B_{\nu}$.}

 Among the sets of parameters that we used in our calculations,
 one of those yielding the closer agreement with the experimental results is:
$J_1=-0.002137$, $J_2=-0.0005$, $J_3=0.002792$, $J_4=0.0010$, $J_5=0.000755$,

$B_1=-0.000482$, $B_2=-0.00023$, $B_3=-0.0001555$, $B_4=-0.00009$, $B_5=-0.000053$, $D_2=0.0005$.

 With these parameters we have at low-temperatures (LT) a conical
 structure with $\theta \sim 60^o$ and $\alpha \sim 114^o$.

 The experimental values are $\theta \sim 58^o$ and $\alpha
 \sim 133^o$ (see ref.2, taking into account that therein $\alpha$
 refers to the rotation of the spin at site ($\frac{1}{2},\frac{1}{2},1$)
 (instead of (0,0,1), as in this work)
 with respect to that at (0,0,0), and that their lattice constant in the $z$
 direction is $c$ while ours is $\frac{1}{2}c$.

 As $T$ increases, both $\theta$ and $\alpha$ increase, until the helical
 structure is attained (with $\alpha \sim 144^o$). This happens at
  $T_{c1} \sim 439$ K. Meanwhile, $\bar{S}$ and $\bar{m{^2}}$ decrease from
 their LT values 1.5 and 2.25.
 Upon a further increase of $T$ (keeping $\theta = 90^o$) $\alpha$
 keeps increasing, reaching $180^o$ (collinear structure) for
 $T_{c2} \sim 448$ K.

 Finally, both $\theta$ and $\alpha$ are kept constant at $90^o$
 and $180^o$ respectively. As $T$ increases,
 $\bar{S}$ and $\bar{m^2}$ decrease until, for $ T = T_{c3} \sim 1041$ K,
 the former gets equal to zero (paramagnetic structure).
 Another set of parameters obtained from the same $X_n$'s is

 $J_1=-0.00223$, $J_2=-0.0005$, $J_3=0.00297$, $J_4=0.0010$,
 $J_5=0.0008$,

 $B_1=-0.000482$, $B_2=-0.00034$, $B_3=-0.0001556$,
 $B_4=-0.0001$, $B_5=-0.000053$, $D_2=0.0005$.

 The results are qualitatively the same as for the previous set, and the $T_c$'s are
 442.5, 455.3 and 1116 K. Notice that while $T_{c1}$ and $T_{c2}$ decrease with
 respect to the previous ones, $T_{c3}$ increases (a behaviour we found in many other cases). 

 Experimentally, the transition "conical $\longleftrightarrow$ helical" is indeed
 observed ( $T_{c1} \sim 320$ K), but from the helical the system
 goes to the PM structure ( $T_{c2} \sim 420$ K) skipping the collinear structure\cite{4}.
 However, the complete sequence of transitions obtained in this work is observed
 in a slightly different compound, namely La$_{1-x}$Y${_x}$Mn$_2$Ge$_2$, for
 $ 0.1 \lesssim x \lesssim 0.2 $ (see fig.13 of ref.4). There, as in our results, the temperature range
 of the helical structure is relatively small.

 It is worthwhile to notice that in the PM phase the spins are not
 totally disordered (i.e. the states with $m = \pm \frac{1}{2}$ are
 less populated than those with $m = \pm \frac{3}{2}$, so that
 $\bar{m{^2}}$ is not equal to 1.25) until a much higher $T$ is reached.
\section{Numerical calculations. LaMn$_2$Si$_2$}

 As for LaMn$_2$Ge$_2$, the interaction constants are obtained from
 the $X_n$'s used in ref.7. 
Concerning the $ S $ of Mn, now both {\em ab initio} calculations
 \cite{7,12} and experimental results\cite{2,5} give a magnetic moment
 near to $\frac{5}{2}\mu_{\rm B} $, which by Land\'e formula and for
 a quenched orbital momentum gives a total angular momentum
 (denoted $S$ in this work) equal to 1.
 In this case no conical structure is obtained for our Hamiltonian.
 However, {\em ab initio} studies\cite{12} indicate that this lowering
 of the magnetic moment is due to Si-Mn hybridization, so that, as for X = Ge,
 we have $S=1.5$.

 Another possibility, leading to the same $S$, is that of having an incomplete quenching, $L=1$,
 with the total spin momentum and the total angular momentum
 both equal to $ \frac{3}{2}$. This would give a Land\'e factor $g = 26/15 $ i.e.
 a magnetic moment $ \mu = 2.6 \mu_B$, in good agreement with the above value.

 Now the set of parameters which we found to give the closest agreement with experiment is:

 $J_1=-0.00145406$, $J_2=-0.0003$, $J_3=0.00177942$, $J_4=0.0010$, $J_5=0.0006$,

 $B_1=-0.00045278$, $B_2=-0.0003$, $B_3=-0.00024431$, $B_4=-0.00017$, $B_5=-0.00010941$, $D_2=0.000$.

 The behaviour of the $J_\nu$, $B_\nu$ in terms of $\nu$ is very similar to that of the Ge compound.

 Proceeding as for the latter we have at LT a conical structure with
 $\theta \sim 53^o$ and $\alpha \sim 139^o$ The agreement of these angles with experiment
 is not as good as for the Ge compound \footnote{This comes from the
 {\em ab initio} results of ref.12, as pointed out therein.}.

 As $T$ increases, both $\theta$ and $\alpha$ increase, but now the former does it more
 slowly, so that $\alpha$ reaches $ 180^o$ (i.e. the canted structure is attained) while
$\theta$ is still nearly $ 54^o$. This happens at $T_{c1} \sim 269$ K.
 
When $T$ (keeps increasing (with $\alpha = 180^o$) $\theta$ increases more rapidly,
 reaching $90^o$ for $T = T_{c2} \sim 393$ K.
 This sequence of transitions is the one observed experimentally \cite{5}.

 Finally, in the collinear structure $\bar{S}$ and $\bar{m^2}$ decrease
  with increasing $T$, until, for $T = T_{c3} \sim 1061$ K, the former
 gets equal to zero (paramagnetic structure). The experimental values reported in ref.5
 are $T_{c1} = 45$ K and $ T_{c2} = 305$ K (the $T > 305$ K region is not studied therein).

\section{About the MFA's.}
 There are many ways of making a MFA in systems like those we study here. 
 One can approximate the exact expression of the energy of the spin at the reference site ($i=0$)
 in the field of the spins at the other sites ($j$) in different ways, and
 thereafter replace $m_j$ and $m_j^2$ by their average values $\bar{S}$ and $\overline{m^2}$) 
 respectively. Some of such approximations are the following:
 
- since in the expression for the total energy the $i$ and $j$ indices are dummy, one can 
 replace the terms with $(m_i^2 + m_j^2)$  in the summation over $i,j$ by
 $ 2 m_i^2 $ or $2 m_j^2$ before of choosing site $i=0$ as reference;

  - one can replace in $\hat{H}_J$ each spin operator $\hat{\vec{S}_i}$ by
 $\vec{\bar{S}}_i + \hat{\vec{\Delta}}_i$
 with $ \hat{\vec{\Delta}}_i \equiv \hat{\vec{S}}_i - \vec{\bar{S}}_i $,
 and neglect higher-order terms in the $\Delta$ operators \cite{8};

 - one can do the same thing for the $m_i m_j$ term in (8);
 
 - for the  $m_i^2 m_j^2$ one can proceed analogously to the case of bilinear terms introducing
 the operators $ \hat{\Delta}_{2;i} \equiv \hat{\vec{S_i} } ^2  - \overline{m_{i}^2} $.

\vspace{2 mm}

 Having chosen the approximate expression of the energy, one can make different choices for the excited
 states available to the spin (see below).
\vspace{2 mm}

 In this work we found that one can obtain completely different results with different MFA's.
 The most interesting example of this (and the only one for which we have
 an explanation!) is the spectacular decrease of the calculated critical temperatures
 when the spin at the reference site is allowed to have a set
 of excited states with a continuous energy spectrum, instead of a discontinuous one.

 In ref.7, with a certain set of parameters and a MFA using a discontinuous spectrum, 
 we obtained for the $ T_c$'s of LaMn$_2$Ge$_2$ the  values 1943 K, 2318 K and 4211 K.
 With the same parameters and one of the MFA's used in this work (continuous spectrum)
 the corresponding values are 468, 473 and 977. A similar situation occurs for
 the Ge compound.
    This difference in the $Tc$'s can be explained as follows.

 In ref.7 the spin at site $i$ could be in any of the eigenstates
 $ \mid m >$ with $ m =  \pm\frac{1}{2}, \pm\frac{3}{2} $ with respect to the $(\theta,\alpha)$ direction
  corresponding to the directions of the other spins. As for any given $T$ $\theta$ and
 $\alpha$, as well as $\bar{S}$ and $\overline{m^2}$), are fixed, for any set of values of these four
 quantities the spin at the reference site needs a finite energy in order to jump to an excited state. 

In the present work, as described in section III, that spin can have (in addition to
 the different $m$ values) any orientation $(\theta_0,\phi_0)$, i.e it can vary its orientation by an
 infinitesimal angle, so that it can acquire a small degree of desorder at a very low $T$.
 This disordering causes a decrease in $\bar{S}$, which in its
 turn reduces the field at the reference site, making easier for the spin
 to jump to an excited state. 
 In addition to this, at higher $T$'s the $m = S-1$ state begins to be populated, so that
 also $\overline{m^2}$ starts to decrease from its LT value $ S^2$.
 This feedback process must be what causes the great decrease of the calculated $ T_c$'s.

 \vspace{2 mm}

 Concerning the way of calculating the energy of the state at the reference site,
 in this work we tried several formulae, i.e. several MFA's, and ended up using
 three of them (that we call MFA0, MFA2 and MFA4).
 Let us consider $ \sum_{i,j}{ }^{'} U_{JB;ij} $
 ( in which, however, as what we need is the energy per site the summation over $i$ is not carried out).
.

 In MFA0 we simply replace $(m_i$ and $m_i^2)$ by $\bar{S}$ and $\overline{m^2}$) 
 respectively.

 In MFA2 we replace $ \sum_{i,j}{ }^{'} B_{ij} (m_i^2 + m_j^2)((p_ij^2 + 1)$ by
 $ 2 \sum_{i,j}{ }^{'} B_{ij} m_i^2((p_ij^2 + 1) $, after which we take $i$ as
 the reference site ($i=0$). This is a usual procedure in order to have a more exact value
 of the total energy before making a MFA. In this case, in which we do not carry out the summation
 over $i$, it is not clear this to be an improvement.

 In MFA4 we use the $\hat{\vec{\Delta}}$ and  $\hat{\Delta}_2$ operators for the terms
 bilinear and biquadratic in $m_i$ and $m_j$, after which we proceed as for the MFA2.

 In all cases the spin at site $i$ is given as described in section III.

 In the calculations we carried out for the Ge compound the sequence 
 "$\!\!$ conical $\longleftrightarrow$ helical $\longleftrightarrow$ collinear $\longleftrightarrow$ PM $\!\!$" 
 was obtained for several sets of parameters with the MFA2, but not with
 the MFA4 (where in most cases the system of equations to be solved (section IV) ceases to have a solution
 above a certain temperature).

 Surprisingly, for the Si compound something nearly opposite is true: the sequence
 " $\!\!$conical $\longleftrightarrow$ canted $\longleftrightarrow$ collinear $\longleftrightarrow$ PM $\!\!$" 
 was obtained for three sets of parameters with the MFA4 and only for one set with the MFA2
 (in the latter case the free energy is discontinuous at $T = T_{c2}$).

 With the MFA0 we obtained the mentioned sequences for only one set of parameters for each of the compounds
 (in the case of Si, with the discontinuity at $T_{c2}$). 

\section{Conclusion.}
 This work is concerned with the hypothesis that the great variety of
 magnetic structures observed in different RT$_2$X$_2$ compounds at
 different temperatures can be explained by assuming that the localized
 magnetic moments are subjected to bilinear and biquadratic interactions
 and that the interaction constants can be derived from {\em ab initio}
 calculations of the energies of different unstable structures.
 The following results of this work give support to this hypothesis:

 - starting from {\em ab initio} energies we find a set of interaction constants
 for which the observed structures
  and the transitions between them are predicted for LaMn$_2$Si$_2$;

 - the interaction constants we obtained by the same procedure for LaMn$_2$Ge$_2$ yield the
  observed structures and the transitions between them of a very
  similar compound, i.e. La$_{0.8}$Y$_{0.2}$Mn$_2$Ge$_2$;

 - for both compounds the dependence of the constants on the interatomic distance is
 perfectly sound for both the bilinear and biquadratic interactions;

 - due to the many competing interactions, the behaviour of the system is
 very sensitive to small changes of the constants, so that it can be
 expected that the same model can be applied to other compounds of
 this family, for which different magnetic structures have been observed.

  If our hypothesis is valid, this paper is only a first step towards a complete theory.
 Among the additional studies that should be carried out we can list the following.
 
 Further work is required to improve the calculation of the interaction constants.

 It is important to understand the pros and cons of each MFA, in order to adopt
 the most adequate one.

 It is likely that more neighbours must be included in the calculation.

 One should also take into account a possible variation of the interaction constants with
 temperature due to the change of the lattice constants, which implies a variation of the
 interatomic distances.

\section{Acknowledgments}

The author is grateful to Dr. A. M. Llois for suggesting this kind of work (the search of a
  spin Hamiltonian making use of {\em ab initio} energy calculations), and to Dr. A.M.Llois,
  Prof. S.Bl\H{u}gel and Dr. S.Di Napoli for useful discussions.

 \end{document}